\documentstyle[12pt,axodraw]{article}

\catcode`@=12
\newcommand{\mrm}[1]{\mbox{\rm #1}}
\newcommand{\beq}{\begin{equation}}
\newcommand{\eeq}{\end{equation}}
\newcommand{\nn}{\nonumber}
\newcommand{\bea}{\begin{eqnarray}}
\newcommand{\eea}{\end{eqnarray}}
\newcommand{\eq}[1]{eq.~(\ref{#1})}
\newcommand{\rfn}[1]{(\ref{#1})}
\newcommand{\Eq}[1]{Eq.~(\ref{#1})}
\newcommand{\db}{\hspace{-0.2ex}\not\hspace{-0.7ex}D\hspace{0.1ex}}
\newcommand{\sla}[1]{\hspace{-0.1ex}\not\hspace{-0.5ex} #1\hspace{0.1ex}}

\newcommand{\ea}{{\it et al.}}

\newcommand{\ie}{{\it i.e. }}
\newcommand{\eg}{{\it e.g. }}

\newcommand{\np}[1]{{ Nucl. Phys. }{\bf #1}}
\newcommand{\pl}[1]{{ Phys. Lett. }{\bf #1}}
\newcommand{\pr}[1]{{ Phys. Rev. }{\bf #1}}

\def\lsigmal{{\cal L}^{\sigma L}}
\def\lsigmar{{\cal L}^{\sigma R}}
\def\lsigma{{\cal L}^{\sigma}}

\def\llr{{\cal L}^{LR}}

\begin{document}
\thispagestyle{empty}
\begin{flushright} 
UCRHEP-T302\\ 
March 2001\
\end{flushright}
\vspace{0.3in}
\begin{center}
{\Large	\bf Enhancement of Radiatively Induced 
Magnetic Moment Form-Factors of Muon:\\
an Effective Lagrangian Approach \\}
\vspace{1.2in}
{\bf Martti Raidal \\}
\vspace{0.2in}
{ \sl National Institute of Chemical Physics and Biophysics, Tallinn 
10143, Estonia, and\\}
\vspace{0.1in}
{\sl Physics Department, University of California, Riverside, 
California 92521, USA\\}
\vspace{1.2in}
\end{center}
\begin{abstract}
Using an effective lagrangian approach, we identify a class of models
in which the loop-induced magnetic moment form-factors of muon are enhanced 
by possibly large factors  
$(\Lambda^2_F/\Lambda^2)(m_\tau/m_\mu)\ln(m_\tau^2/\Lambda^2)$ or
$(\Lambda^2_F/\Lambda^2)\ln(m_\mu^2/\Lambda^2),$ 
where $\Lambda$ is the scale of new physics and $\Lambda_F$ is the Fermi 
scale. These follow from left- and right-chirality mixing dimension-8 
operators which for relatively small $\Lambda,$ 
as required to explain the new $(g_\mu-2)$ measurement,
dominate over dimension-6 operators.
Thus significant enhancement of new physics contributions to 
$(g_\mu-2)$ and, in the presence of intergenerational couplings, 
also to the $\mu\to e\gamma$ decay rate is possible.
We discuss the compatibility of the $(g_\mu-2)$ and $\mu\to e\gamma$ 
experimental data in this case and comment on the enhancement of the 
electron anomalous magnetic moment. An explicit model is presented to 
illustrate the general results. 
\end{abstract}
\newpage
\baselineskip 24pt

\section{Introduction}

The recently announced measurement \cite{bnl} of the muon anomalous 
magnetic moment:
\begin{equation}
a_\mu^{exp} = {g_\mu - 2 \over 2} = 116592020 (160) \times 10^{-11},
\label{amu}
\end{equation}
which differs from the standard-model (SM) prediction \cite{czma} by 
$2.6\sigma$:
\begin{equation}
\Delta a_\mu = a_\mu^{exp} - a_\mu^{SM} = 426 \pm 165 \times 10^{-11},
\label{adev}
\end{equation}
indicates that a relatively large positive new contribution to $a_\mu$ 
is needed, hinting thus at new physics above the electroweak scale.  
If the new contribution to $(g_\mu-2)$ is induced at loop level, 
which is the case in models of neutrino masses \cite{meie1},
supersymmetric models \cite{esusy}, 
models with extra dimensions \cite{eext}, 
models with enlarged gauge \cite{egauge}, Higgs \cite{ehiggs} or 
fermion \cite{eferm} 
sectors, models with leptoquarks \cite{eleptoq} or models of compositness
\cite{ecomp}, then the largeness of 
$\Delta a_\mu$ implies usually a quite strong uper bound of order 
${\cal O}(100)$ GeV on the new physics mass scale. While in some models
there exist a mechanism to enhance the  new contribution to $a_\mu$
(\eg, in supersymmetric models it is enhanced by large value of 
$\tan\beta$), in others the masses and couplings of new particles 
should be tuned to satisfy the experimental value of $\Delta a_\mu.$

At the same time, the new physics which gives rise to $\Delta a_\mu$
should likely affect also other leptonic observables. The most sensitive
of them are the lepton flavour violating (LFV) processes, such as
the decay $\mu\to e \gamma,$  which occurs in the presence of flavour 
non-diagonal couplings. The connection between $\Delta a_\mu$ and 
$\mu\to e \gamma$ is particularly natural because both of them 
are induced by the same type of magnetic operators. In addition,
if the same new physics gives also rise to non-zero neutrino masses
and mixings, the LFV processes must be large \cite{meie1} 
due to almost maximal mixings in the neutrino sector \cite{sol}.

To classify the models above,  an effective lagrangian description of
new physics is a useful tool \cite{eff}. For $\Delta a_\mu$ the 
effective lagrangian analyses was done in Ref. \cite{ew} in which
all dimension-6 operators inducing $\Delta a_\mu$  were considered.
Their general result agrees qualitatively with the results in each
specific model: the new physics scale $\Lambda$ should be relatively
low, just above the Fermi scale  $\Lambda_F.$ On the one hand, this 
implies that  
the effective lagrangian description is not suitable for precision
calculations because the higher order operators (dimension-8 and higher)
may not be suppressed compared to the dimension-6 operators
and in some cases may even dominate. On the other hand, because of the 
conceptual simplicity, the effective lagrangian language is still 
useful to understand the generic behaviour of certain class of models
under consideration. Once the general properties are understood,
the precision calculations may be performed in each model separately.
This is the philosophy we adopt in this paper.

The purpose of this paper is to identify a class of models in 
which left- and right-chirality mixing 
dimension-8 effective operators inducing $\Delta a_\mu$ at one loop
dominate over the dimension-6 operators. These operators were not 
considered in Ref. \cite{ew}. The factors  
$(\Lambda^2_F/\Lambda^2)(m_\tau/m_\mu)\ln(m_\tau^2/\Lambda^2)$
and $(\Lambda^2_F/\Lambda^2)\ln(m_\mu^2/\Lambda^2)$ occuring
in the magnetic moment form-factors in these models may be large enough
to significantly enhance the new physics contributions to  $\Delta a_\mu$
and  to $\mu\to e \gamma.$ We analyze these two processes here, and 
comment also on the $m_\tau/m_e$ enhancement of the electron anomalous 
magnetic moment and its connection to $\Delta a_\mu.$
To illustrate the general result we propose a simple model 
in which the enhancement occurs, and perform exact calculations in
that model.

\section{Effective lagrangian approach}

Assuming the new physics to appear at the scale $\Lambda,$
the relevant terms in the effective lagrangian contributing directly 
to magnetic moments are
\begin{equation}
\label{eq:sigmal}
\lsigmal=
\frac{\alpha^{\sigma L}_{ij}}{(4\pi)^2 \Lambda^2} e\ \overline{e_{iL}}
\sigma_{\mu\nu} i \db e_{jL} F^{\mu\nu} + \mathrm{h.c.}\,,
\end{equation}
and
\begin{equation}
\label{eq:sigmar}
\lsigmar=
\frac{\alpha^{\sigma R}_{ij}}{(4\pi)^2 \Lambda^2} e\ \overline{e_{iR}}
\sigma_{\mu\nu} i \db e_{jR} F^{\mu\nu} + \mathrm{h.c.}\,.
\end{equation}
Here  $e_{iL}$ and $e_{iR}$ are chiral charged-lepton fields,
$\db = \sla{\partial} + i e\sla{A},$ the Lorentz indices are
$\mu,\nu,$ and the indices $i,j$ denote generations. 
Since these terms cannot be obtained from renormalizable vertices 
at tree level, we expect them to be generated at one loop. 
That is the reason we already included a factor $(4\pi)^2$ in the denominator. 

We have written $\lsigmal$ and $\lsigmar$ in a particular form involving only
left or right chiral fields. The two operators
could be combined by using the equations of motion for the leptons.
In this case we obtain 
\begin{equation}
\lsigma =
\frac{1}{(4\pi)^2 \Lambda^2} e\
\overline{e_L} \sigma_{\mu\nu} F^{\mu\nu} \left(\alpha^{\sigma L} m_e+
m_e \alpha^{\sigma R}\right) e_R  + \mathrm{h.c.}\,,
\label{lsigma}
\end{equation}
where where $m_e$ is the charged lepton mass matrix and the generation
indices are suppressed. 
In chiral theories, like the ones we want to consider, magnetic moments appear
always in the form \eq{lsigma} and are proportional to the fermion
masses. In more general theories with
chirality explicitly broken independently of the fermion masses, operators
like \eq{lsigma} but with an arbitrary matrix $M$ could arise.

The dimension-6 four fermion operators inducing \eq{lsigma} at one loop
level are considered in Ref. \cite{ew}. Here we consider the effective
lagrangians of the type
\bea
\label{lr}
\llr&=&
\frac{\alpha^{LR}_{ik,lj}}{\Lambda^4} (\tilde\Phi^+\Phi)
(\overline{e_{iL}} e_{kL}^c)(\overline{e_{lR}^c} e_{jR}) +\mathrm{h.c.}\,,
\eea
where $\Phi$ is the SM Higgs doublet, $\tilde\Phi=\tau_2\Phi^*$ and
$e_{L,R}^c = (e_{L,R})^c$ are the charge conjugated fields.
The couplings $\alpha^{LR}_{ik;lj}$ 
are symmetric with respect to the exchanges $i\leftrightarrow k$ and
$l \leftrightarrow j.$ However, the pairs of indices $(ik)$ and $(lj)$
are totally independent. Therefore there is no need to write down the 
corresponding $RL$ operator which is of the form of the hermitian
conjugate of \eq{lr}; all such independent terms are already included in  
\eq{lr}. We assume $\alpha^{LR}_{ik;lj}$ to  be real.
We have chosen the form \eq{lr} because it is simple for loop calculations. 
One could perform a Fierz transformation of \eq{lr} to get rid of
the charge conjugate fields. However, the simplicity will be lost in this
case and  the tensor operators will occur. In addition, the
operators of type \eq{lr} arise naturally in the class of models 
we shall consider.

\begin{figure}
\begin{center}
\begin{picture}(270,110)(0,0)
\ArrowLine(35,10)(135,10)
\ArrowLine(135,10)(235,10)
\ArrowArc(135,50)(40,90,270)
\ArrowArc(135,50)(40,270,90)
\Photon(170,70)(240,110){4}{5}
\GCirc(135,10){3}{0}
\Text(135,0)[]{$\Lambda$}
\Text(50,0)[]{$ l^-_{i}$}
\Text(220,0)[]{$ l^-_{j}$}
\Text(135,105)[]{$l^-_{k} $}
\Text(135,80)[]{$m_{k} $}
\Text(135,90)[]{$\times $}
\end{picture}
\end{center}
\caption{\it The enhanced one loop contribution to magnetic form-factors. 
The chirality flip in the internal lepton line is denoted by $\times.$}
\label{diag1}
\end{figure}
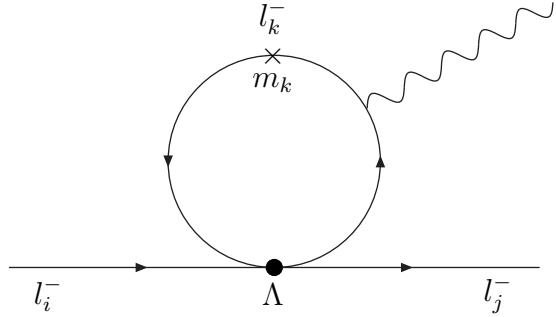

After the Higgs boson will acquire the vacuum expectation value (vev) $v,$
the corresponding four-fermion operator will occur which gives a
contribution to  magnetic moments via the one loop diagram depicted in
Fig. \ref{diag1}. On the one hand, this is suppressed by 
$v^2/\Lambda^2$ compared to the similar dimension-6 four-fermion operator 
contributions. On the other hand, however, 
because of the left and right mixing chiral structure of 
\eq{lr}, the helicity flip must occur now in the internal fermion line
as explicitly noted in Fig. \ref{diag1}. This will imply the possible
enhancement\footnote{In SUSY models the enhancement of $\Delta a_\mu$ has
a similar origin \cite{esusy}: 
for large $\tan\beta$ the slepton $LR$ mass terms 
$m_{LR}$ become large. However, no large logarithms occur in SUSY models
since the superpartner masses are of the same order of magnitude.}  
factors $m_k/m_i$ compared to the usual case
when the chirality flip occurs in the external line. In addition,
since the fermion masses $m_k$ are very small compared to the scale
$\Lambda,$ one expects large logarithms $\ln (m_k^2/\Lambda^2)$ to
occur in the magnetic moment form-factors. 
Again, these large logarithms  never occur 
when the chirality flip occurs in the external line.
Numerically, for example for $\Lambda\sim 1$ TeV, $k=\tau,$ $i=\mu,$
the factor $|(m_\tau/m_\mu)\ln (m_\tau^2/\Lambda^2)|\approx 210$
overcomes the suppression factor $v^2/\Lambda^2\approx 1/16,$
and significant enhancement of loop induced magnetic moment 
form-factors will occur.

To show this with an explicit calculation, we express the relevant
matrix element in the most general way as
\bea
{\cal M} &=& e {\bar u}(p_j) \left[ \;
\left (f_{E0} + \gamma_5 f_{M0}\right) \gamma_{\nu}  \left (
g^{\lambda \nu} - \frac {q^{\lambda} q^{\nu}}{q^2}\right )
\right. \nn \\
&  & +
\left.
(f_{M1} + \gamma_5 f_{E1})\; i\; \sigma^{\lambda \nu} \frac{q_{\nu}}{m_{i}}
\right] u(p_{i}) \epsilon_\lambda(q)\,,
\label{j1}
\eea    
where $p_i,\,\,p_j$ are the lepton momenta and $q$ is the momentum of the
photon. We calculate the form factors $f_{E0}$, $f_{E1}$, $f_{M0}$ 
and $f_{M1}$ induced only by \eq{lr} via the diagram in
Fig. \ref{diag1} (similar diagrams without the enhancement are considered 
in Ref. \cite{rs}). 
For  $i$ denoting the initial and  $j$ the final 
state particle and $k$ the particle running in the loop the answer reads 
\bea
&&f_{M0}=f_{E0}=0\,, \\
&&f_{M1}=\sum_k
\frac{2 \left( \alpha^{LR}_{jk,ki}+ \alpha^{LR}_{ik,kj} \right)}{(4\pi)^2}
\frac{m_i^2}{\Lambda^2}\frac{v^2}{\Lambda^2} \frac{m_k}{m_i} 
F\left( \frac{m_k^2}{\Lambda^2};\frac{-q^2}{\Lambda^2} \right)\,, 
\label{m1}\\
&&f_{E1}=\sum_k
\frac{2 \left( \alpha^{LR}_{jk,ki}- \alpha^{LR}_{ik,kj} \right)}{(4\pi)^2}
\frac{m_i^2}{\Lambda^2}\frac{v^2}{\Lambda^2} \frac{m_k}{m_i} 
F\left( \frac{m_k^2}{\Lambda^2};\frac{-q^2}{\Lambda^2} \right)\,,
\label{e1}
\eea 
where
\bea
F\left(x;y \right)=2-\ln x + \sqrt{1+\frac{4 x}{y}} 
\ln\left( \frac{\sqrt{4 x+y}+\sqrt{y}}{\sqrt{4x+y}-\sqrt{y}} \right)\,,
\eea
and we have taken $q^2\le 0.$ In the limit of on-shell photon, which 
is the case we are interested in, the function $F$ simplifies to
\bea
\lim_{y\to 0}F\left(x;y \right)= 4 - \ln x\,.
\eea
It is clear from \eq{m1},\rfn{e1} that an enhancement of the 
form-factors by $m_k/m_i\ln(m_k^2/\Lambda^2)$ will occur. This is 
the case for both anomalous magnetic moments with $i=j$ as well as for
the transition magnetic moments with $i\neq j.$

Let us now turn to studies of specific observables.
The new physics contribution from the effective lagrangians \eq{lsigma} 
and \eq{lr} to the muon anomalous magnetic moment is  given by
\bea
\Delta a_\mu=\frac{2 m^2_\mu}{(4\pi)^2\Lambda^2}
\left(\alpha^{\sigma L}_{\mu\mu}+\,\alpha^{\sigma R}_{\mu\mu} +
\sum_k 4 \alpha^{LR}_{\mu k,k\mu} \frac{v^2}{\Lambda^2}\frac{m_k}{m_\mu}
\left[ 4-\ln \frac{m^2_k}{\Lambda^2} \right]\right)\,.
\label{eq:sigmav}
\eea
As we expect $\alpha^{\sigma L}\sim \alpha^{\sigma R}\sim \alpha^{LR}$
the loop induced contribution from \eq{lr} clearly dominates.

Similarly,
the $l_i \to l_j \gamma$ rate divided by the $l_i \to l_j \nu_i \bar \nu_j$
rate is given by
\bea
R(l_i\to l_j\gamma)=\frac{96 \pi^3\alpha}{G_F^2m_{l_i}^4}
\left( |f_{M1}|^2 + |f_{E1}|^2 \right) \,,
\eea
where $\alpha=1/137$ and $G_F$ is the Fermi constant. For the decay
$\mu\to e \gamma$ one has $i=\mu,$  $j=e$ and the corresponding 
transition form-factors are 
\bea
f_{M1}&=&\frac{ m^2_\mu}{(4\pi)^2\Lambda^2}
\left(\alpha^{\sigma L}_{e\mu}+\,\alpha^{\sigma R}_{e\mu} +
\sum_k 2\left( \alpha^{LR}_{e k,k \mu}+\alpha^{LR}_{\mu k,k e} \right)
\frac{v^2}{\Lambda^2}\frac{m_k}{m_\mu}
\left[ 4-\ln \frac{m^2_k}{\Lambda^2} \right]\right)\,, \\
f_{E1}&=&\frac{ m^2_\mu}{(4\pi)^2\Lambda^2}
\left(\alpha^{\sigma L}_{e\mu}-\,\alpha^{\sigma R}_{e\mu} + 
\sum_k 2\left( \alpha^{LR}_{e k,k \mu}-\alpha^{LR}_{\mu k,k e} \right)
\frac{v^2}{\Lambda^2}\frac{m_k}{m_\mu}
\left[ 4-\ln \frac{m^2_k}{\Lambda^2} \right]\right)\,.
\eea
Again, a significant enhancement of the $\mu\to e \gamma$ rate is
expected.

Let us now turn to discussion of our general results.
Assuming that only one of the couplings 
$\alpha^{LR}_{\mu\tau,\tau\mu}\sim 4\pi$
or $\alpha^{LR}_{\mu\mu,\mu\mu}\sim 4\pi$ 
is non-zero at the time while the other 
vanishes, the 90\% confidence-level experimental result 
$\Delta a_\mu\ge 215\cdot 10^{-11}$ \cite{czma}
implies upper bounds on the new physics scale $\Lambda$ as shown in Table~1.
Due to the enhancement the bounds are of order ${\cal O}(1)$ TeV 
rather that of order ${\cal O}(100)$ GeV as expected in models with no 
enhancement. If the off-diagonal couplings are large too, 
$\alpha^{LR}_{e k,k \mu}\sim\alpha^{LR}_{\mu k,k e}\sim 4\pi,$ 
this would imply
\bea
R(\mu\to e\gamma)\ge 1.5 \cdot 10^{-3}\,,
\eea
which is orders of magnitude above the present limit
$R(\mu\to e\gamma)< 1.2 \cdot 10^{-11}$ \cite{mue}. This shows the
correlation between $\Delta a_\mu$ and $R(\mu\to e\gamma)$ in
these models. Because the experimental value of $\Delta a_\mu$ 
fixes $\Lambda$ to be at relatively low scale, and because
the process $\mu\to e\gamma$ is much more sensitive to new physics
than $(g_\mu-2),$ the only way to suppress the $\mu\to e\gamma$
rate is to suppress the LFV couplings. The upper bounds on the  
couplings $\alpha^{LR}_{\mu k,k e}$ obtained for the present
limit on $\mu\to e\gamma$ as well as for the expected limit 
$R(\mu\to e\gamma)< 2 \cdot 10^{-14}$ \cite{psi} are presented 
in Table 1. Here we have assumed that $\alpha^{LR}_{e k,k \mu}=
\alpha^{LR}_{\mu k,k e}.$
It follows that the LFV couplings should be smaller than
at least $10^{-3}$ in order not to go into conflict with the 
experimental data.

\begin{table}
\begin{tabular}{|c|c||c|c|}
\hline\hline
\multicolumn{2}{||c||}{$\Delta a_\mu \ge215\cdot 10^{-11}$}
 & $R(\mu\to e\gamma)< 1.2\cdot 10^{-11}$ &
$R(\mu\to e\gamma)<2\cdot 10^{-14}$  \\
\hline\hline
$\;\alpha^{LR}_{\mu\tau,\tau\mu}\sim 4\pi\;$&$\;\Lambda\le 2.82\;$ TeV $\;$& 
$\alpha^{LR}_{\mu\tau,\tau e}\le 1.1\cdot 10^{-3}$  & 
$\alpha^{LR}_{\mu\tau,\tau e}\le 4.7\cdot 10^{-5}$ \\
\hline
$\alpha^{LR}_{\mu\mu,\mu\mu}\sim 4\pi$ & $\Lambda\le 1.47\;$ TeV & 
$\alpha^{LR}_{\mu\mu,\mu e}\le 1.1\cdot 10^{-3}$  & 
$\alpha^{LR}_{\mu\mu,\mu e}\le 4.7\cdot 10^{-5}$ \\
\hline\hline
\end{tabular}
\caption{\it Upper bounds on the scale $\Lambda$ if either internal
$\tau$ or $\mu$ contribution to $\Delta a_\mu$ dominates. The upper bounds
on the LFV couplings $\alpha^{LR}$ are given for the same 
$\Lambda.$ 
}
\end{table}

Let us now consider the anomalous magnetic moment of electron in our 
scenario. Because the necessary chirality flip occurs in the internal 
fermion line, also the new contribution to the electron anomalous magnetic
moment  $\Delta a_e$ is proportional to the tau mass $m_\tau$ rather than to 
the electron mass $m_e.$ This implies the enhancement factor 
$(\Lambda^2_F/\Lambda^2)(m_\tau/m_e)\ln(m_\tau^2/\Lambda^2)$ 
compared to the usual case. Therefore, if 
$\alpha^{LR}_{e \tau,\tau e}\approx \alpha^{LR}_{\mu \tau, \tau \mu}$
then \eq{adev} automatically implies
$\Delta a_e \approx (m_e/m_\mu)\Delta a_\mu \sim 10^{-11}$
in our scenario. This is more than an order of magnitude larger than the
current experimental uncertainty on $a_e.$ 
Therefore, either 
$\alpha^{LR}_{e \tau,\tau e} < \alpha^{LR}_{\mu \tau, \tau \mu},$
or one must reconsider the SM contributions to $a_e$ 
(and also the quantities derived from it, such as $\alpha_{QED}$).

Finally, two comments are in order. First, 
the origin of the enhancement of the
magnetic moment form-factors discussed here
is in the left-right chiral structure
of the effective lagrangian \eq{lr}. One can easily construct similar
lagrangians by replacing two of the leptons by some very heavy
exotic leptons $E$, for example. In this case the enhancement factor
$m_E/m_\mu$ from the chirality flip in the loop is huge. However, 
this type of models are beyond our considerations here.
Second, if the error bars in $\Delta a_\mu$ will be increased, the 
bounds in Table 1 should be revised.

\section{An explicit model}   

Here we present and explicit model in which the operators of type 
\eq{lr} occur. The Higgs sector of the model consists of the usual 
SM doublet with the $SU(2)_L\times U(1)_Y$ quantum numbers
\bea
\Phi =\left(\begin{array}{c} \phi^+ \\ \phi^0 \end{array}\right)\sim (2,1/2)\,,
\eea
and two additional scalar fields, a triplet $\xi$ and a singlet $\chi$:
\bea
\xi =\left(\begin{array}{cc} \xi^+/\sqrt{2} & \xi^{++}\\ 
 -\xi^0 & -\xi^+/\sqrt{2} \end{array}\right)\sim (3,1)\,,
~~~~~~~~~\chi=\chi^{++}\sim (1,2)\,.
\eea
The latter two fields carry lepton number $-2.$
The triplet couples to lepton doublets $L$ via the Yukawa interaction
\bea
{\cal L}_\xi= f_{ij} L_i^T C^{-1}\, i\tau_2\, \xi\, L_j + \mathrm{h.c.}\,,
\label{yukt}
\eea
while the lagrangian for the $\chi$
coupling to lepton singlets is 
\begin{equation}
{\cal L}_\chi = h_{ij} \overline{e_{iR}^c} e_{jR}\,\chi^{++} +
\mathrm{h.c.}\,.
\label{yuks}
\end{equation}              
Here the Yukawa coupling matrices $f_{ij},h_{ij}$ are symmetric in the 
generation indices $i,j$.
We assume them to be real.

The most general Higgs potential containing these fields is
\begin{eqnarray}
V &=& m_0^2 \,\Phi^\dagger \Phi + m_\xi^2\, \mrm{Tr}[\xi^\dagger \xi] + 
 m_\chi^2\,\chi^\dagger \chi +
{1 \over 2} \lambda_1 (\Phi^\dagger \Phi)^2 + 
{1 \over 2} \lambda_2 \mrm{Tr}[\xi^\dagger \xi]^2 + 
{1 \over 2} \lambda_3 (\chi^\dagger \chi)^2 + 
 \nonumber \\ 
&& \lambda_4 \mrm{Tr}[\xi^\dagger \xi^\dagger]\mrm{Tr}[\xi \xi] +
\lambda_5 (\Phi^\dagger \Phi)\mrm{Tr}[\xi^\dagger \xi]
+ \lambda_6 \Phi^\dagger \xi^\dagger \xi \Phi +
\lambda_7 (\chi^\dagger \chi)\mrm{Tr}[\xi^\dagger \xi] +
\lambda_8 (\chi^\dagger \chi)(\Phi^\dagger \Phi) + \nn \\
&& \lambda_9 \left( \chi \Phi^\dagger \xi^\dagger \tilde\Phi + h.c. \right) +
\left( {\mu \over \sqrt 2} 
\Phi^\dagger \xi \tilde \Phi + h.c. \right) \,,
\label{V}
\end{eqnarray}
where $m_0^2 < 0$, but $m_\xi^2 > 0$. The neutral components of the 
fields acquire vevs as $\phi^0\to\phi^0+v/\sqrt{2},$
$\xi^0\to\xi^0+u/\sqrt{2}.$ Thus the $SU(2)_L$ gauge symmetry is
broken as in the SM but, because $m_\xi^2 > 0,$ the lepton number 
is not broken spontaneously. Thus there is no Majoron in this model.
This is the biggest difference between the model presented here and
the original Gelmini-Roncadelli model \cite{gero}. 
Nevertheless, neutrinos may have
Majorana masses in this model because lepton number is broken 
explicitly by the last dimensionful term in \eq{V}. This follows 
 from the first derivative minimization conditions 
\begin{eqnarray}
m_0^2 - \mu u + {1 \over 2} \lambda_1 v^2 + {1 \over 2} \lambda_5 u^2 
&=& 0, \nn\\ 
m_\xi^2 u  - {1 \over 2} \mu v^2 + {1 \over 2} \lambda_2 u^3 + 
{1 \over 2} \lambda_5 u v^2  &=& 0.
\label{mincon}
\end{eqnarray}
Therefore, $v^2 \simeq -2m_0^2/\lambda_1$ as usual, but $u \simeq  \mu
 v^2/2m_\xi^2$, with $u \ll v$. 
The small vev $u$ of the triplet, 
which gives masses to the neutrinos via \Eq{yukt}, is proportional to the 
value of $\mu$ and inversely proportional to the square of the Higgs triplet 
mass, \ie  $m_\xi^2.$  Thus the smallness of the neutrino mass must follow
from the smallness of lepton number breaking parameter $\mu$ which
can be achieved, \eg, in models with extra dimensions \cite{marasa}.
This model has also a number of unique experimental signatures at
future collider experiments \cite{cr}.

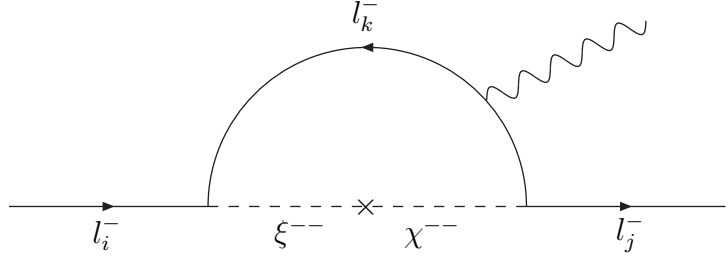
\begin{figure}
\begin{center}
\begin{picture}(270,110)(0,0)
\ArrowLine(0,10)(75,10)
\ArrowLine(195,10)(270,10)
\DashLine(75,10)(195,10){4}
\ArrowArc(135,10)(60,0,180)
\Photon(180,50)(240,80){4}{5}
\Text(110,0)[]{$\xi^{--}$}
\Text(160,0)[]{$\chi^{--}$}
\Text(37,0)[]{$ l^-_{i}$}
\Text(235,0)[]{$ l^-_{j}$}
\Text(135,83)[]{$l^-_k$}
\Text(135,10)[]{$\times $}
\end{picture}
\end{center}
\caption{\it Diagrams giving rise to enhanced $\Delta a_\mu$ and 
$l_i \to l_j \gamma$ in our example model. The 
photon can be attached to any charged line.}
\label{diag2}
\end{figure}

For our studies here only the doubly charged Higgs bosons are important.
The enhanced contribution to $\Delta a_\mu$ and $\mu\to e\gamma$
arises from the diagrams depicted in Fig. \ref{diag2}.
In order to get the same chiral structure as in the effective lagrangian
\eq{lr}, mixing of the left-handedly (triplet) and right-handedly 
(singlet) interacting  particles must occur. This is explicitly shown in 
Fig. \ref{diag2}. The doubly charged Higgs boson 
mass matrix following from \eq{V} is in the basis $(\xi^{++},\chi^{++})$  
given by
\begin{equation}
{\cal M}^2 = \frac{1}{2}\left[ \begin{array}{cc} 
\left(\lambda_6 + \mu/u\right) v^2 + 4 \lambda_4 u^2 &
\lambda_9 v^2 \\
\lambda_9 v^2 & 2 m^2_\chi +  \lambda_8 v^2 +\lambda_7 u^2
\end{array} \right]\,.
\label{m++}
\end{equation}
Notice that the off-diagonal entry is proportional to $\sim v^2.$ 
Thus the mixing between the two doubly charged Higgses 
is roughly given by $\sim v^2/m^2_\chi.$ 
This is nothing but the extra suppression factor $ v^2/\Lambda^2$  
appearing in \eq{lr} after the gauge symmetry breaking. 
In this context the requirement of going to 
higher order (dimension-8) operators is explained by the requirement
of having the left-right mixings of the Higgs bosons.

Explicit calculation shows that the enhanced new physics contribution to 
$a_\mu$ in our model is 
\bea
\Delta a_\mu=
\sum_k \frac{f_{\mu k} h_{k\mu}}{4\pi^2} \frac{m_k}{m_\mu} \sin 2\theta 
\sum_a (-1)^{1+a}\frac{m^2_\mu}{M^2_a}
\left[ \frac{7}{2}-\ln \frac{m^2_k}{M_a^2} \right] \,,
\label{a}
\eea
where the second sum over $a=1,2$ goes over the two
doubly charged Higgs boson mass eigenstates,
and the angle $\theta$ is the mixing angle between them.
$\theta$ can be calculated from \eq{m++}.

Similarly, the transition form-factors for the decay $\mu\to e\gamma$ are
\bea
f_{M1}&=&
\sum_k \frac{\left( h_{\mu k} f_{e k}+ f_{\mu k} h_{ke} \right)}{(4\pi)^2}
\frac{m_k}{m_\mu} \sin 2\theta 
\sum_a (-1)^{1+a}\frac{m^2_\mu}{M^2_a}
\left[ \frac{7}{2}-\ln \frac{m^2_k}{M_a^2} \right]
\,, \\
f_{E1}&=&
\sum_k \frac{\left( h_{\mu k} f_{e k}- f_{\mu k} h_{ke} \right)}{(4\pi)^2}
\frac{m_k}{m_\mu} \sin 2\theta 
\sum_a (-1)^{1+a}\frac{m^2_\mu}{M^2_a}
\left[ \frac{7}{2}-\ln \frac{m^2_k}{M_a^2} \right]
\,.
\eea
Notice the two differences compared to the effective lagrangian form 
factors. First and the less important one is that the numerical factor $7/2$
appears in the square brackets rather than factor 4.  This is because
we have taken into account all  contributing diagrams in our model;
of course, the dominant leading logarithm is the same here
as in the effective lagrangian case. Secondly, there are two Higgs mass
eigenstates contributing; the effective lagrangian result will be 
approximately achieved only if $M_2\gg M_1\sim \Lambda.$ 
The exact result depends on the values of the parameters in \eq{m++}.
For a numerical example we take $f_{\mu\tau}h_{\tau\mu}=1,$ all $\lambda=1,$ 
$(\mu/u)v^2=m^2_\chi$ and $u\ll v$ in \eq{m++}. Then 
$\Delta a_\mu > 215\cdot 10^{-11}$ implies $m_\chi<1.4$ TeV. 
Therefore we emphasize that exact calculations in each particular model
are important to explain quantitatively the observed $\Delta a_\mu.$

Let us now discuss the decay $\mu\to e\gamma.$  One of the motivations
to consider this model here is that it gives an enhanced $\Delta a_\mu$
as well as small Majorana neutrino masses at the same time. Both of
them are experimentally observed quantities. Our knowledge
of the neutrino mass matrix implies that at least $f_{\mu\tau}$ ,
and possibly also $f_{\mu e}$ entries in the Yukawa matrix $f$ must
be large. This is because of almost maximal mixing angles in the
neutrino sector \cite{sol}. To satisfy the experimental constraints on 
$R(\mu\to e\gamma)$ one has to suppress the couplings 
$h_{\mu\tau} f_{e \tau}$ and $f_{\mu \tau} h_{\tau e}.$
Since this is impossible for $f_{\mu \tau}$ if we want to induce
the observed neutrino properties, we conclude that 
the LFV couplings of $h_{ij}$ must be very much suppressed
(see Table~1).

\section{Conclusions}

We have shown that the left-right chirality mixing 
dimension-8 effective operators of
type \eq{lr} give radiatively induced
contributions to muon magnetic moment form-factors which are
enhanced by 
$(\Lambda^2_F/\Lambda^2)(m_\tau/m_\mu)\ln(m_\tau^2/\Lambda^2)$
compared to the dimension-6 operator contributions. As the measured 
$\Delta a_\mu$ requires the scale $\Lambda$ to be of order TeV, these
dimension-8 operator contributions dominate over the dimension-6 ones,
implying enhancement of new physics contributions to $a_\mu$ and to 
the decay $\mu\to e\gamma.$ Using  effective lagrangians
we have derived constraints on $\Lambda$ from the observed $\Delta a_\mu$
as well as constraints on the LFV couplings from $\mu\to e\gamma.$
We have illustrated this general result by an explicit model
with a scalar triplet and a singlet. This model is motivated by the fact
that it can generate the observed neutrino masses and the
enhanced $\Delta a_\mu$ at the same time.
Because the new physics scale $\Lambda$ is low, we emphasize the 
importance of exact calculations in each particular model. 

In this scenario also  the new contribution to the anomalous magnetic 
moment of  electron is proportional to the internal lepton mass and
thus enhanced by 
$(\Lambda^2_F/\Lambda^2)(m_\tau/m_e)\ln(m_\tau^2/\Lambda^2).$
Numerically this exceeds the present experimental uncertainty on 
$\Delta a_e$ and requires
further suppression of the $e\tau$ couplings.

\vspace*{0.5cm}

{\bf Acknowledgement.} 
I thank Ernest Ma, Arcadi Santamaria and Jose Wudka for discussions.
This work was supported in part by the U.~S.~Department of Energy
under Grant No.~DE-FG03-94ER40837 and by the Estonian Science Foundation 
Grant No. 3832.

\newpage
\bibliographystyle{unsrt}

\end{document}